\documentclass[10pt,journal,final]{IEEEtran}
\IEEEoverridecommandlockouts

\newtheorem{remark}{Remark}
\usepackage{amsfonts,amssymb}
\usepackage{array}
\usepackage{ifpdf}
\usepackage{cite}
\usepackage{hyperref}
\usepackage{url}
\usepackage{stfloats}
\usepackage{bm}
\usepackage{verbatim}
\usepackage{color,xcolor}
\usepackage[caption=false,font=footnotesize,labelfont=rm,textfont=rm]{subfig}
\usepackage{cite}
\usepackage{subfloat}

\DeclareSubrefFormat{parens}{#1(#2)}

\ifCLASSINFOpdf
\usepackage[pdftex]{graphicx}
\graphicspath{{../pdf/}{../jpeg/}}
\DeclareGraphicsExtensions{.pdf,.jpeg,.png}
\else
\usepackage[dvips]{graphicx}

\graphicspath{{../eps/}}

\DeclareGraphicsExtensions{.eps}
\fi

\usepackage{amsmath, amssymb}

\usepackage{algorithm}
\usepackage{algorithmic}

\usepackage{makecell}

\newenvironment{proof}{{\indent\indent\it Proof:}}{\hfill $\square$\par}
\allowdisplaybreaks[4]

\begin{document}
	\captionsetup[figure]{name={Fig.},font=footnotesize,labelfont=rm,labelsep=period,singlelinecheck=off} 
	\title{Beamforming Design for Intelligent Reflecting Surface Aided Near-Field THz Communications }
	
	\author{Chi~Qiu, Qingqing~Wu, Wen~Chen, Meng~Hua, Wanming~Hao, Mengnan~Jian, Fen~Hou
		\thanks{
			C. Qiu and F. Hou are with the State Key Laboratory of Internet of Things for Smart
			City and the Department of Electrical and Computer Engineering, University
			of Macau, Macau SAR, 999078, China
			(e-mail: chi.qiu77@connect.um.edu.mo; fenhou@um.edu.mo).
			Q. Wu and W. Chen are with the Department of Electronic Engineering, Shanghai
			Jiao Tong University, Shanghai 201210, China (e-mail: qingqingwu@sjtu.edu.cn; wenchen@sjtu.edu.cn).
			M. Hua is with
			the Department of Electrical and Electronic Engineering, Imperial College
			London, London SW7 2AZ, UK (e-mail: m.hua@imperial.ac.uk).
			W. Hao is with the School of Electrical and Information Engineering, Zhengzhou University, Zhengzhou 450001, China (e-mail: iewmhao@zzu.edu.cn).
			 M. Jian is with ZTE Corporation, Shenzhen 518057, China (e-mail:
			jian.mengnan@zte.com.cn).
	}\vspace{-0.6cm}}
	\maketitle
	\begin{abstract}
		Intelligent reflecting surfaces (IRSs) operating in the terahertz (THz) band have gained considerable interest due to their high spectral bandwidth. With large-scale IRS deployment, transceivers are likely to be in the near-field region, where the near-field beam split effect becomes a major challenge for wideband IRS beamforming. This effect causes beam deviation from the intended location, leading to significant gain loss and waste in spectrum resource. While delay-based IRS has emerged as a potential solution, current beamforming schemes generally assume unbounded range time delays (TDs).  In this paper, we first investigate the near-filed beam split issue at the IRS. Then, we extend the piecewise far-field model to the IRS, based on which, a double-layer delta-delay (DLDD) IRS beamforming scheme is proposed. Specifically, we employ an element-grouping strategy and the TD imposed on each sub-surface of IRS is achieved by a series of TD modules. This method significantly reduces the required range of TDs.In addition, a sub-optimal joint beamforming scheme is proposed to address the double-stage beam split at the base station and the IRS.
		Numerical results show that the proposed schemes can effectively mitigate the near-field beam split and achieve near-optimal performance.
	\end{abstract}
	\begin{IEEEkeywords}
		IRS, THz, near-field, joint beamforming, and beam split.
	\end{IEEEkeywords}
	
	\vspace{-0.4cm}
	\section{Introduction}
	Terahertz (THz) communication is a promising technique for future sixth-generation (6G) wireless communications due to its ultra-wide bandwidth,  which is expected to achieve data rates of up to terabits per second (Tb/s) theoretically \cite{THz}. However, THz signals face significant challenges, including severe transmission attenuation and poor scattering, which limits the transmission range. Fortunately, intelligent reflecting surface (IRS) is a promising solution to these challenges. By using a large number of passive reflection elements, an IRS can steer the signals to any desired direction \cite{IRS2}. This capability allows the IRS to enhance signal strength, making it a practical solution for future THz communications \cite{IRS_THz}.
	
	Although significant efforts have been made, most existing IRS beamforming schemes do not perform well when the signal bandwidth is large, especially in practical THz communication systems. This is primarily because practical IRS is usually equipped with frequency-independent phase-shifting circuits \cite{IRS1}, resulting in IRS-aided communications only being able to achieve frequency-independent beamforming. This leads to the beam split issue in the IRS-aided THz system. In \cite{DAM}, a delay-adjustable metasurface was proposed, where each IRS element is connected to a time-delay (TD) module. The beam split effect could be addressed because the imposed delay enabled frequency-dependent
	IRS beamforming. However, since the number of IRS elements required in THz scenarios is relatively huge, equipping each element with a TD module becomes impractical, leading to excessive hardware cost and power consumption. In both \cite{time_delay} and \cite{wideband_src}, the delay-based IRS was designed in a sub-connected manner. The above works only addressed the far-field beam split effect. In fact, due to the high propagation losses and the short wavelength, a THz IRS is expected to consist
	of a massive number of passive reflecting elements, making the transceivers very likely located in the near-field region of the IRS. Although \cite{far_near}  proposed a delay-based solution to overcome the near-field beam split effect at the IRS, the required number of TD modules was still large and it also ignored the practical limitation of TD technology in the THz band. In fact, research on TD realization in the THz band is still in its early stages, with a limited achievable delay range. Specifically, the TD module discussed in \cite{TDrange} was only able to achieve a maximum delay range of $14.272$ picoseconds (ps). Given this limitation, it is crucial to take into account the practical THz TD capabilities when designing the delay-based IRS. 
	
 Considering the above, we aim to investigate the mitigation of near-field beam split effect with practical short-range TD modules in an IRS-aided THz communication system in this paper.    The main contributions are summarized as follows. First, we extend the piecewise far-field model in \cite{piece_wise} to the uniform planar array (UPA)-structured IRS.
	By dividing the IRS into multiple sub-surfaces, the base station (BS) and the user can be reasonably assumed to be in the far-field range of each small sub-surface. Such segmentation enables us to decompose the phase difference among different IRS elements that causes beam split into two separate components, i.e., the inter-surface counterpart and the intra-surface counterpart. Second, motivated by the delta-delay-phase precoding proposed in \cite{delta}, we propose a novel double-layer delta-delay (DLDD) architecture to alleviate the inter-surface phase discrepancy. Specifically, the first layer TD network takes care of the phase discrepancy along one axis of IRS, while the second layer takes care of that along the other axis. Third, the discussions are extended from single-antenna to multi-antenna system, where the double-stage beam split at both the BS and the IRS occurs. To combat with such effect, we propose a sub-optimal joint beamforming scheme. Finally, numerical results reveal the necessity of delay-enabled frequency-dependent IRS beamforming and demonstrate the effectiveness of the proposed DLDD architecture in mitigating the near-field beam split using practical short-range TD modules. Furthermore, it is shown that the proposed joint beamforming scheme  achieves enhanced achievable rate performance with strong robustness in a wide bandwidth range.
	
	\begin{figure}[t]
		\centering
		\includegraphics[width=2.6in]{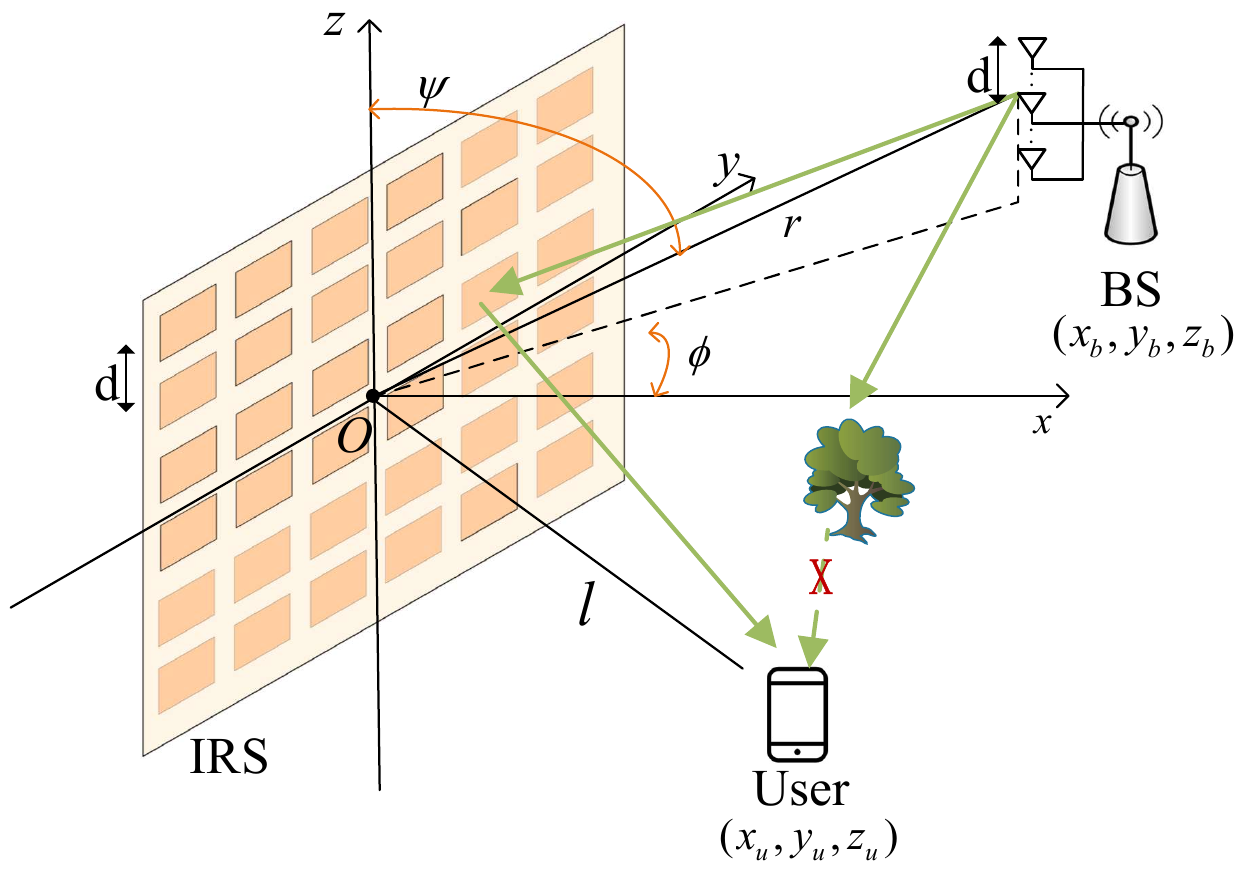}
		\vspace{-0.2cm}
		\caption{An IRS-aided THz communication system.}
		\label{sysmod}
		\vspace{-0.6cm}
	\end{figure}
	
	\vspace{-0.3cm}
	\section{System Model}	
	We consider an IRS-aided THz communication system, where an IRS is deployed to assist the downlink transmission from the BS to a user.\footnote{ The design can be straightforwardly extended to the scenarios with multi-antenna user. Also, it can be applied to the scenario with the direct path by modifying the equivalent channel model accordingly. } As shown in Fig. \ref{sysmod}, suppose the BS is equipped with an $N_t$-element uniform linear array along the $z$-axis and the $N$-element UPA-structured IRS is located on the $y-z$ plane, both with antenna/elements separation of $d$.  The center of BS is located at $(x_{\text{b}},y_{\text{b}},z_{\text{b}})$, and the coordinate of the $n_t$-th element is then given by $(x_{\text{b}},y_{\text{b}},z_{\text{b}}-\Delta_{n_t}^{N_t}d)$, where $\Delta_{a}^{b}=a-1-\frac{b-1}{2}$.  The IRS is centered at the origin, with $N_y$ elements along the $y$-axis and $N_z$ elements along the $z$-axis, satisfying $N=N_y\times N_z$.  Then, the coordinate of the $(n_y,n_z)$-th IRS element, with $n_y=\mathcal{N}_y\triangleq\{1,\dots,N_y\}$ and $n_z=\mathcal{N}_z\triangleq\{1\dots,N_z\}$, is given by $(0,\Delta_{n_y}^{N_y}d,\Delta_{n_z}^{N_z}d)$. The coordinate of the single-antenna user is assumed to be $(x_{\text{u}},y_{\text{u}},z_{\text{u}})$. Both of the BS and the user are considered to be within the Fraunhofer distance of the IRS, i.e., $R =\frac{2D^2}{\lambda_c}$, where $D$ is the maximum aperture of the IRS and $\lambda_c$ denotes the corresponding signal wavelength.
	 The center frequency is denoted by $f_c$ and the bandwidth is denoted by $B$, which is equally divided into $M$ subcarriers.  Then, each subcarrier frequency can be denoted by $f_m=f_c+\frac{B}{M}\left(m-1-\frac{M-1}{2}\right)$, with $m=\mathcal{M}\triangleq\{1,\dots,M\}$.
	
	Assuming that the direct link between the BS and the user is blocked by obstructions, the user can only receive signals that are reflected by the IRS. In the THz band, the power gains associated with scattering paths are notably lower compared to the direct line-of-sight (LoS) path, thus we only focus on the LoS channel \cite{wideband_src}.\footnote{ It is assumed the channels are perfectly estimated using the pilot-assisted methods \cite{CSI}.}
	The LoS near-field channel between the BS and the IRS  at the $m$-th subcarrier is denoted by $\mathbf{G}_m=\left[\operatorname{vec}\left(\mathbf{g}_m^{1}\right),\cdots,\operatorname{vec}\left(\mathbf{g}_m^{N_t}\right)\right]\in\mathbb{C}^{N\times N_t}$, where  the entries of $\mathbf{g}_m^{n_t}$ are given by
	\begin{equation}
		\left[\mathbf{g}_m^{n_t}\right]_{n_y,n_z}=\frac{\alpha_{m}}{r^{n_t}_{n_y,n_z}} e^{-j2\pi\frac{f_m}{c}r^{n_t}_{n_y,n_z}},	
	\end{equation}
where $\alpha_{m}\triangleq \frac{c}{4\pi f_m}$, $\frac{\alpha_{m}}{r^{n_t}_{n_y,n_z}}$ denotes the free-space path loss, and $r^{n_t}_{n_y,n_z}$ denotes the distance between the $n_t$-th BS antenna and the $(n_y,n_z)$-th IRS element, which can be calculated as  $r^{n_t}_{n_y,n_z}=\left(x_{\text{b}}^2+\left(y_{\text{b}}-\Delta_{n_y}^{N_y}d\right)^2+\left(z_{\text{b}}-\Delta_{n_z}^{N_z}d\right)^2\right)^{\frac{1}{2}}$.
	
	Similarly, the LoS near-field channel between the IRS and the user at the $m$-th subcarrier is denoted as $\mathbf{h}_m\in\mathbb{C}^{N\times1}$, whose entries are given by \begin{equation}	\vspace{-0.2cm} 	\left[\mathbf{h}_m\right]_{n_y,n_z}=\frac{\alpha_{m}}{l_{n_y,n_z}}e^{-j2\pi\frac{f_m}{c}l_{n_y,n_z}},
	\end{equation}where  $l_{n_y,n_z}=\left(x_{\text{u}}^2+\left(y_{\text{u}}-\Delta_{n_y}^{N_y}d\right)^2+\left(z_{\text{u}}-\Delta_{n_z}^{N_z}d\right)^2\right)^{\frac{1}{2}}\hspace{-0.2cm}$ represents the  distance between the $(n_y,n_z)$-th IRS element and the user. The IRS reflection coefficients matrix is expressed as\begin{equation}
		\boldsymbol{\Theta}=\operatorname{diag}(\boldsymbol{\theta})=\operatorname{diag}\left(\left[e^{j \theta_1}, e^{j \theta_2}, \ldots, e^{j \theta_N}\right]^T\right),
	\end{equation}where $\theta_{n} \in [0,2\pi), n\in\{1,\dots,N\}$, is the phase shift of the $n$-th IRS reflecting element \cite{IRS1}. Then, the achievable rate per subcarrier can be calculated as\begin{equation}
	R_m=\log_2\left(1+\frac{P_m\left|\mathbf{h}_m^T\boldsymbol{\Theta}\mathbf{G}_m\mathbf{w}_m\right|^2}{\sigma_m^2}\right),
	\end{equation} where $P_m$ is the BS transmit power, $\mathbf{w}_{m}$ is the BS beamforming vector, and $\sigma_m^2$ is the noise power at the $m$-th subcarrier.

	\section{Wideband IRS Beamforming Scheme}
	In this section, the near-field beam split effect at the IRS is analyzed. Then, the piecewise far-field channel model for UPA-structured IRS is derived, based on which, a wideband IRS beamforming scheme based on DLDD architecture is proposed to address the beam split problem. The number of BS antenna is reduced to $N_t=1$ to simplify expression, while the scenario with multi-antenna BS will be discussed in Section IV. Thus, the BS-IRS channel is simplified to $\mathbf{g}_m$.
	\subsection{Near-field Beam Split Effect at the IRS}
\begin{figure}[!t]
		\centering
		\subfloat[Near-field beam split by conventional beamforming.\label{beamsplit}]{\includegraphics[width=1.7in]{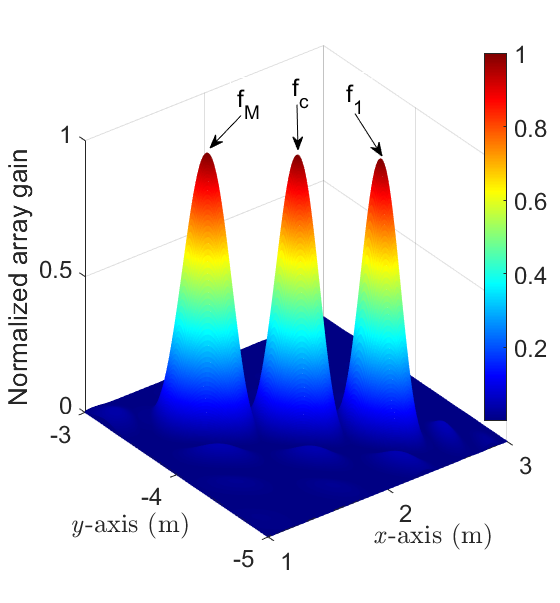}}%
		\hfil
		\subfloat[The beams generated by the proposed beamforming design.\label{proposed_beamsplit}]{\includegraphics[width=1.7in]{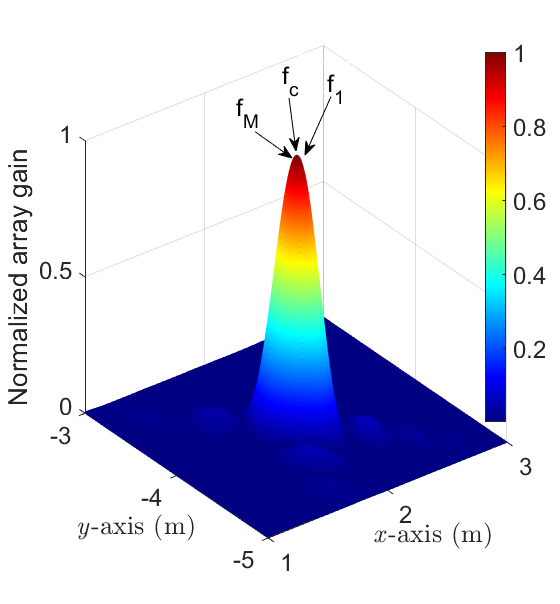}}%
		
		\caption{The IRS reflected beams under different cases.}
		\label{fig}
		\vspace{-0.5cm}
	\end{figure}

	We define the array gain of the IRS at the $m$-th subcarrier as\begin{subequations}\vspace{-0.2cm}\begin{align}
				\hspace{-0.2cm}\eta\left(f_m\right)&\!=\!\left|\mathbf{h}_m^T\boldsymbol{\Theta}\mathbf{g}_m\right|\\
				&\!=\!\left|\sum_{n_y=1}^{N_y}\sum_{n_z=1}^{N_z}\!e^{-j2\pi \frac{f_m}{c}\left(r_{n_y,n_z}\!+l_{n_y,n_z}\right)} \!e^{j\theta_{n_y,n_z}}\right|.
			\end{align}
	\end{subequations}
	The conventional narrowband IRS beamforming design aims at generating the reflected beams towards the target location. We adopt such an IRS beamforming vector based on $f_c$, expressed as \begin{equation}\label{NB_BF}
		\theta_{n_y,n_z}=\frac{2\pi}{\lambda_c}\left(r_{n_y,n_z}+l_{n_y,n_z}\right), n_y\in\mathcal{N}_y,n_z\in\mathcal{N}_z.
	\end{equation} Owing to the frequency-independent characteristic of \eqref{NB_BF}, the generated beams cannot perfectly point towards the target location in the whole bandwidth, which leads to severe gain loss. In Fig. \ref{fig}\subref{beamsplit}, we illustrate this by plotting  the normalized array gain at $f_c$ and two edge subcarriers, $f_1$ and $f_M$, achieved by \eqref{NB_BF}, under the same setup that will be specified in Section IV. It can be observed that the near-field beam split causes the beams at different subcarriers to split towards different locations. This restricts the user to receive signals that are only around the center frequency.
	

	\vspace{-0.2cm}
	\subsection{Piecewise Far-Field Channel Modeling for UPA IRS}
	To reduce hardware cost, we adopt the sub-connected structure. Specifically, we divide the IRS into $K$ sub-surfaces, with $K_y$ sub-surfaces along the $y$-axis and $K_z$ sub-surfaces along the $z$-axis, i.e., $K=K_y\times K_z$. It is assumed that each sub-surface contains the adjacent $S\times S$ elements that share a common delay, satisfying $S=N_y/K_y=N_z/K_z$. The coordinate of the  center of $(k_y,k_z)$-th sub-surface, with $k_y\in\mathcal{K}_y\triangleq\{1,\dots,K_y\}$ and $ k_z\in\mathcal{K}_z\triangleq\{1,\dots,K_z\}$, is then expressed as $\left(0,\Delta_{k_y}^{K_y}Sd,\Delta_{k_z}^{K_z}Sd\right)$. Inspired by the piecewise far-field channel model in \cite{piece_wise}, we next derive the channel model for the considered UPA-structured IRS. Specifically, by dividing the entire IRS into multiple sub-surfaces, each of which contains much fewer elements compared to the original surface. This partition significantly reduces the near-field range for each sub-surface. As a result, even if the BS/user is within the near-field region of the entire IRS, it can be reasonably assumed to be in the far-field region of each individual sub-surface. Let $r$ be the distance between the center of IRS and the BS, which can be written as
	$r=\sqrt{x_{\text{b}}^2+y_{\text{b}}^2+z_{\text{b}}^2}$. Then, the distance between the BS and the center of $(k_y,k_z)$-th sub-surface can be written as \begin{subequations}
		\begin{align}
			\hspace{-0.5cm}r_{k_y,k_z}=&\sqrt{x_{\text{b}}^2+\left(y_{\text{b}}-\Delta_{k_y}^{K_y}Sd\right)^2+\left(z_{\text{b}}-\Delta_{k_z}^{K_z}Sd\right)^2}\\
			=&\left(r^2+\left(\Delta_{k_y}^{K_y}Sd\right)^2-2\Delta_{k_y}^{K_y}Sdr\sin\left(\psi\right)\sin\left(\phi\right)\nonumber \right.\\ 
			&\left.+\left(\Delta_{k_z}^{K_z}Sd\right)^2-2\Delta_{k_z}^{K_z}Sdr\cos\left(\psi\right)\right)^{\frac{1}{2}},
		\end{align}
	\end{subequations}where $\phi$ and $\psi$ are the azimuth and elevation angles of the BS with respect to (w.r.t.) the center of the IRS, respectively, as illustrated in Fig. \ref{sysmod}. 
	To this end, the distance $r^{k_y,k_z}_{s_y,s_z}$ between the BS and the $(s_y,s_z)$-th element, with $s_y,s_z\in\mathcal{S}\triangleq\{1,\dots,S\}$, of the $(k_y,k_z)$-th sub-surface  can be expressed as \begin{subequations}
		\begin{align}\label{eq7}
			r^{k_y,k_z}_{s_y,s_z}=&\left(r^{\,2}_{k_y,k_z}+\left(\Delta^S_{s_z}d\right)^2-2\Delta^S_{s_z}dr_{k_y,k_z}\cos\left(\psi_{k_y,k_z}\right)\nonumber\right.\\ 
			&\hspace{-1.4cm}\left.+\left(\Delta^S_{s_y}d\right)^2-2\Delta^S_{s_y}dr_{k_y,k_z}\sin\left(\psi_{k_y,k_z}\right) \sin\left(\phi_{k_y,k_z}\right)\right)^{\frac{1}{2}} \\
			&\hspace{-1.4cm}\stackrel{(a)}{\approx} \!r_{k_y,k_z}\!\!-\!\Delta^S_{s_z}d\cos\!\left(\psi_{k_y,k_z}\!\right)\!-\!\Delta_{s_y}^Sd\sin\!\left(\psi_{k_y,k_z}\!\right)\!\sin\!\left(\phi_{k_y,k_z}\!\right)\!,
		\end{align}
	\end{subequations}where   $\phi_{k_y,k_z}$ and  $\psi_{k_y,k_z}$ are the azimuth and elevation angles of the BS w.r.t. the center of $(k_y,k_z)$-th IRS sub-surface, respectively, satisfying\vspace{-0.2cm}\begin{equation}\label{angle1}
		\sin\left(\phi_{k_y,k_z}\right)=\frac{y_{\text{b}}-\Delta_{k_y}^{K_y}Sd}{\left(x_{\text{b}}^2+\left(y_{\text{b}}-\Delta_{k_y}^{K_y}Sd\right)^2\right)^{\frac{1}{2}}},
	\end{equation}\vspace{-0.2cm} \begin{equation}\vspace{-0.2cm}\label{angle2}
		\sin\left(\psi_{k_y,k_z}\right)=\frac{\left(x_{\text{b}}^2+\left(y_{\text{b}}-\Delta_{k_y}^{K_y}Sd\right)^2\right)^{\frac{1}{2}}}{r_{k_y,k_z}},
	\end{equation} and \begin{equation}\vspace{-0.2cm}\label{angle3}
		\cos\left(\psi_{k_y,k_z}\right)=\frac{z_{\text{b}}-\Delta_{k_z}^{K_z}Sd}{r_{k_y,k_z}}.\vspace{0.2cm}
	\end{equation} $(a)$ in \eqref{eq7} holds due to the first-order Taylor expansion $(1 + x)^{\frac{1}{2}} \approx 1+ \frac{1}{2} x$ and the ignorance of $\left(\Delta^S_{s_y}d\right)^2$ and $\left(\Delta^S_{s_z}d\right)^2$.\footnote{Such ignorance arises from the fact that the distance between the  $r_{k_y,k_z}$ is generally much larger than the element separations, $\Delta^S_{s_y}d$ and $\Delta^S_{s_z}d$, within a sub-surface.}
	Accordingly, the normalized near-field channel model between the BS and the $(s_y,s_z)$-th IRS element in the $(k_y,k_z)$-th sub-surface is approximated as $\tilde{\mathbf{g}}_m$, whose entries are given by\vspace{-0.5cm} \begin{align}\vspace{-2cm}\left[\tilde{\mathbf{g}}_m\right]^{k_y,k_z}_{s_y,s_z}=&e^{-j2\pi \frac{f_m}{c}r_{k_y,k_z}}e^{j2\pi\frac{f_m}{c}\Delta^S_{s_z}d\cos\left(\psi_{k_y,k_z}\right)}\nonumber\\
		&\times e^{j2\pi \frac{f_m}{c}\Delta_{s_y}^Sd\sin\left(\psi_{k_y,k_z}\right)\sin\left(\phi_{k_y,k_z}\right)}.\vspace{-0.5cm}
	\end{align} 
	Similarly, the normalized near-field channel between the IRS and the user can be approximated as $\tilde{\mathbf{h}}_m$, whose entries are given by\vspace{-0.2cm}\begin{align}\vspace{-0.3cm}
		\left[\tilde{\mathbf{h}}_m\right]^{k_y,k_z}_{s_y,s_z}=&e^{-j2\pi \frac{f_m}{c}l_{k_y,k_z}} e^{j2\pi\frac{f_m}{c}\Delta^S_{s_z}d\cos\left(\upsilon_{k_y,k_z}\right)}\nonumber\\
		&\times e^{j2\pi \frac{f_m}{c}\Delta_{s_y}^Sd\sin\left(\upsilon_{k_y,k_z}\right)\sin\left(\omega_{k_y,k_z}\right)},\vspace{-0.2cm}
	\end{align} where $l_{k_y,k_z}$ is the distance between the user and the center of $(k_y,k_z)$-th sub-surface, while $\omega_{k_y,k_z}$ and $\upsilon_{k_y,k_z}$ are the corresponding azimuth and elevation angles, respectively, defined similarly as \eqref{angle1}, \eqref{angle2} and \eqref{angle3}.
	
	\vspace{-0.2cm}
	\subsection{Proposed DLDD IRS Beamforming}
	
	\begin{figure}[t]
		\centering
		\includegraphics[width=3in]{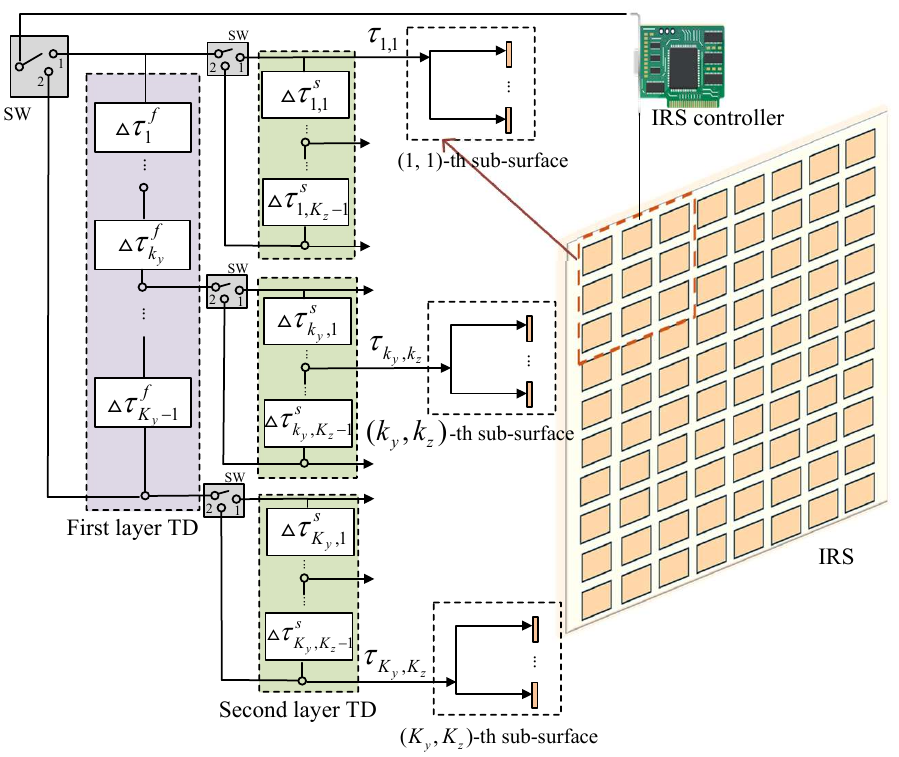}
		\vspace{-0.3cm}
		\caption{The proposed DLDD IRS beamforming architecture.}
		\label{double_layer_TTD}
		\vspace{-0.5cm}
	\end{figure}
	Accordingly, the normalized piecewise far-field cascaded BS-IRS-user channel between the BS and the user, through the reflection of the $(s_y,s_z)$-th IRS element in the $(k_y,k_z)$-th sub-surface, can be denoted by $\tilde{\mathbf{b}}_m$, whose entries are given by \vspace{-0.3cm}\begin{align}\vspace{-0.3cm}\label{cascaded_channel}
		\left[\tilde{\mathbf{b}}_m\right]^{k_y,k_z}_{s_y,s_z}=&e^{-j2\pi \frac{f_m}{c}\left(r_{k_y,k_z}+l_{k_y,k_z}\right)}\nonumber \\
		&\times e^{j2\pi \frac{f_m}{c}\left(\varphi^{k_y,k_z}_{s_y,s_z}+\varepsilon^{k_y,k_z}_{s_y,s_z}\right)},
	\end{align}where $	\varphi^{k_y,k_z}_{s_y,s_z}\!=\!\Delta_{s_y}^Sd\sin\!\left(\!\psi_{k_y,k_z}\!\right)\!\sin\!\left(\!\phi_{k_y,k_z}\!\right)\!+\!\Delta^S_{s_z}d\cos\!\left(\!\psi_{k_y,k_z}\!\right)\!$ and $
		\varepsilon^{k_y,k_z}_{s_y,s_z} \!=\! \Delta_{s_y}^Sd\!\sin\!\left(\!\upsilon_{k_y,k_z}\!\right)\!\sin\!\left(\!\omega_{k_y,k_z}\!\right)\!
		+\!\Delta^S_{s_z}d\cos\!\left(\!\upsilon_{k_y,k_z}\!\right)\!$.
	The piecewise far-field channel model makes it straightforward to decouple the phase in \eqref{cascaded_channel} into two components: the inter-subsurface phase difference $\delta_{k_y,k_z}^r=r_{k_y,k_z}+l_{k_y,k_z}$ among different sub-surfaces, and the intra-subsurface phase difference $\delta^{k_y,k_z}_{s_y,s_z}=\varphi^{k_y,k_z}_{s_y,s_z}+\varepsilon^{k_y,k_z}_{s_y,s_z}$ within each sub-surface. Considering the fact that each sub-surface contains much fewer elements, this reduces the effect of beam split. Hence, we can reasonably deduce that the intra-surface phase difference $\delta^{k_y,k_z}_{s_y,s_z}$ has little effect on the near-field beam split. Thus, we design the TD values to compensate for the inter-surface phase difference $\delta_{k_y,k_z}^r$ in order to alleviate the near-field beam split effect. Accordingly, the required phase shift at the $(k_y,k_z)$-th sub-surface to combat beam split effect at the $m$-th subcarrier is given by\vspace{-0.2cm}\begin{equation}\label{required_phase}
		\vartheta^m_{k_y,k_z}=2\pi f_m\frac{\delta_{k_y,k_z}^r}{c}.
	\end{equation} Consequently, the required delay value is derived by matching the phase shift obtained from a TD module connected to the $(k_y,k_z)$-th sub-surface to that in \eqref{required_phase} as follows\begin{equation}\label{required_TD}
		\tau_{k_y,k_z}=-\frac{\delta_{k_y,k_z}^r}{c},k_y\in\mathcal{K}_y,k_z\in\mathcal{K}_z.
	\end{equation}However, the delay value can be large in this case. For example, with $\delta_{k_y,k_z}^r=3$ meters (m), $\tau_{k_y,k_z}$ can be $1\times 10^4$ ps. To avoid excess TD module's limitation, we propose a DLDD architecture as shown in Fig. \ref{double_layer_TTD}, which utilizes a two-layer network configuration. The first layer TD network consists of $\left(K_y-1\right)$ TD modules arranged in a step-wise manner, generating $K_y$ cumulative delays. Each of these cumulative delays is then fed into an array of $\left(K_z-1\right)$ TD modules. As a result, the second layer consists of a total of $K_y\left(K_z-1\right)$ TD modules. Therefore, the proposed DLDD architecture requires a total of $K_{\text{T}}=\left(K_y-1\right)+K_y\times \left(K_z-1\right)=\left(K-1\right)$ TD modules.
	
	The first-layer TD network is specifically designed to eliminate the phase difference between adjacent sub-surfaces along the $y$-axis. Specifically, the $k_y$-th delay value in this layer, $k_y\in\mathcal{K}_y'\triangleq\{1,\dots,K_y-1\}$, is designed to eliminate the phase difference between the $(k_y,1)$-th and the $(k_y+1,1)$-th sub-surfaces, which is given by\begin{equation}\label{tau1}
		\hspace{-0.35cm}\Delta\tau_{k_y}^{\text{f}}\!=\!\tau_{k_y+1,1}\!-\!\tau_{k_y,1}\!=\!-\frac{\delta_{k_y+1,1}^r\!-\!\delta_{k_y,1}^r}{c}, k_y\in\mathcal{K}_y'.
	\end{equation}Similarly, the second-layer TD network aims at eliminating the phase difference between adjacent sub-surfaces along the $z$-axis. Specifically, the $k_z$-th delay value, $k_z\in\mathcal{K}_z'\triangleq\{1,\dots,K_z-1\}$, of the $k_y$-th group is designed to eliminate the phase difference between the $(k_y,k_z+1)$-th and the $(k_y,k_z)$-th sub-surfaces, which is given by \begin{align}\label{tau2}
		\Delta\tau_{k_y,k_z}^{s}&=\tau_{k_y,k_z+1}-\tau_{k_y,k_z}\nonumber\\
		&=-\frac{\delta_{k_y,k_z+1}^r-\delta_{k_y,k_z}^r}{c}, k_y\in\mathcal{K}_y,k_z\in\mathcal{K}_z'.
	\end{align}Thus, the total delay applied on the $(k_y,k_z)$-th sub-surface is achieved through a cumulative sum of TD values from both layers, expressed as $t_{k_y,k_z}=\sum_{k=1}^{k_y-1}\Delta\tau_{k}^{\text{f}}+\sum_{i=1}^{k_z-1}\Delta\tau_{k_y,i}^{\text{s}}$.
	
	 Such cumulative strategy allows a more efficient utilization of TD modules, as the architecture does not rely on dedicated TD modules for each delay value, significantly alleviating the maximum delay constraints faced by conventional systems \cite{time_delay}. To represent the design mathematically, the parameters of the TD modules in the first layer network is expressed as $\boldsymbol{\tau}_{\text{f}}=[\Delta\tau_{1}^{\text{f}},\dots,\Delta\tau_{K_y-1}^{\text{f}}]$, and that in  the second layer TD network is expressed as $\boldsymbol{\tau}_{s}=[\Delta\tau_{1,1}^{\text{s}},\Delta\tau_{1,2}^{\text{s}}\dots,	\Delta\tau_{K_y,K_z-1}^{\text{s}}]$. As demonstrated in \eqref{tau1} and \eqref{tau2}, the TD values can be either positive or negative. However, TD modules are unable to generate negative delays. Fortunately, we have the following Lemma.

		\textbf{Lemma 1.} Define $\epsilon\left(k_y\right) \triangleq -\frac{\left(y_{\text{b}}-\Delta_{k_y}^{K_y}Sd\right)}{r_{k_y,1}}-\frac{\left(y_{\text{u}}-\Delta_{k_y}^{K_y}Sd\right)}{l_{k_y,1}}$. TD values $\Delta\tau_{k_y}^{\text{f}}$ are strictly positive, i.e., $\Delta\tau_{k_y}^{\text{f}}>0,\forall k_y\in\mathcal{K}_y'$, if $\epsilon\left(K_y\right)<0$. Conversely, TD values are strictly negative, i.e., $\Delta\tau_{k_y}^{\text{f}}<0,\forall k_y\in\mathcal{K}_y'$, if $\epsilon\left(1\right)>0$.
		
		\begin{proof} 
			The first-order derivative of $\delta_{k_y,1}^r$ w.r.t. $k_y$ is given by $\frac{d (\delta_{k_y,1}^r)}{d \left(k_y\right)} =\epsilon\left(k_y\right)$. In addition, since the second-order derivative of  $\delta_{k_y,1}^r$ w.r.t. $k_y$ satisfies $\frac{d^2 (\delta_{k_y,1}^r)}{d \left({k_y}\right)^2} =\frac{x_{\text{b}}^2+\left(z_{\text{b}}-\Delta_{1}^{K_z}Sd\right)^2}{r_{k_y,1}}+\frac{x_{\text{u}}^2+\left(z_{\text{u}}-\Delta_{1}^{K_z}Sd\right)^2}{l_{k_y,1}}\!>\!0, \forall k_y$, it indicates the monotonic increase in $\epsilon_{k_y}$. If $\epsilon\left(K_y\right)<0$, then $\epsilon\left(k_y\right)<0,\forall k_y\leq K_y$, which readily indicates that $\Delta\tau_{k_y}^{\text{f}}\!\!>\!\!0,\forall k_y\in\mathcal{K}_y'$. Conversely, if $\epsilon\left(1\right)>0$, then $\epsilon\left(k_y\right)>0,\forall k_y\geq 1$, implying $\Delta\tau_{k_y}^{\text{f}}\!\!<\!\!0,\forall k_y\in\mathcal{K}_y'$. The proof is thus complete. 
	\end{proof}


	\begin{remark}\label{lemma1}
			The property of $\Delta\tau_{k_y,k_z}^{s}$ can be similarly proved. Thus, we focus on the discussion on $\Delta\tau_{k_y}^{\text{f}}$, which can be directly applied to  $\Delta\tau_{k_y,k_z}^{s}$.
	\end{remark}
	
	Lemma \ref{lemma1} establishes a connection between the property of the TD sequence $\Delta\tau_{k_y}^{\text{f}}$ and the locations of the transceivers. Since the size of the IRS is generally far less than the distances $r_{k_y,1}$ and $l_{k_y,1}$, we reasonably assume that $\epsilon_{k_y}>0$ for all $k_y\in\mathcal{K}_y$ if $\epsilon_{i}>0$ for any $i\in\mathcal{K}_y$, and vice versa. In other words, it is reasonable to assume that $\epsilon_{1}>0$ and $\epsilon_{K_y}<0$ never occur simultaneously. Under such assumption, the signs of TD values $\Delta\tau_{k_y}^{\text{f}},\forall k_y\in\mathcal{K}_y'$ are consistent.   To this end,  we propose to insert a $2$-output switch before each group of TD modules. The switch closes its $1^{st}$ output  when the elements of $\boldsymbol{\tau}_{\text{f}}$ or $\boldsymbol{\tau}_{s}$ are positive, or rather directs the signal through the $2^{nd}$ output to the TD network. Specifically, it is modeled as a binary selecting vector $\mathbf{s}_{\text{\text{f}}}$, given by\begin{equation}
			\mathbf{s}_{\text{\text{f}}}=\frac{1}{2|\Delta\tau_{1}^{\text{f}}|}\left[\begin{array}{l}
				|\Delta\tau_{1}^{\text{f}}|+ \Delta\tau_{1}^{\text{f}}\\
				|\Delta\tau_{1}^{\text{f}}|-\Delta\tau_{1}^{\text{f}}
			\end{array}\right].
	\end{equation}Denote $\mathbf{t}^+_{\text{\text{f}}}=\left[0,\Delta\tau_{1}^{\text{f}},\sum_{k_y=1}^{2}\Delta\tau_{k_y}^{\text{f}}\cdots,\sum_{k_y=1}^{K_y-1}\Delta\tau_{k_y}^{\text{f}}\right]$ and $\mathbf{t}^-_{\text{f}}=\left[\sum_{k_y=1}^{K_y-1}\Delta\tau_{k_y}^{\text{f}},\sum_{k_y=2}^{K_y-1}\Delta\tau_{k_y}^{\text{f}},\cdots,\Delta\tau_{K_y-1}^{f},0\right]$. The resulting delay of the first layer TD network is then given by\vspace{-0.2cm}\begin{equation}
	\mathbf{t}_{\text{f}}=\left[\mathbf{t}^+_{\text{f}},\mathbf{t}^-_{\text{f}}\right]\mathbf{s}_{\text{f}}.\vspace{-0.2cm}
\end{equation}

	\vspace{-0.3cm}
	
	On the other hand, the IRS phase shift is designed to align with the intra-surface phase discrepancy $\delta^{k_y,k_z}_{s_y,s_z}$. Specifically, for the $(s_y,s_z)$-th IRS element in the $(k_y,k_z)$-th sub-surface, we have\begin{equation}\label{IRS}
		\theta^{k_y,k_z}_{s_y,s_z}=\operatorname{mod}\left(-2\pi\frac{f_c}{c}\delta^{k_y,k_z}_{s_y,s_z},2\pi\right),\forall k_y,k_z,s_y,s_z,
	\end{equation} where $\operatorname{mod}$ is the modulo operation. Denote $t_{\text{f},k_y}$ and $t_{\text{s},k_y,k_z}$ as the corresponding entries of $\mathbf{t}_{\text{f}}$ and $\mathbf{t}_{\text{s},k_y}$, where $\mathbf{t}_{\text{s},k_y}$ is the resulting delay of the $k_y$-th group in second layer TD network, derived similarly as above. Thus, the total delay applied on the $(k_y,k_z)$-th sub-surface  is re-expressed as $t_{k_y,k_z}=t_{\text{f},k_y}+t_{\text{s},k_y,k_z}$. With the above design, the array gain is expressed as \eqref{arraygain}, shown at the top of the page. \begin{figure*}[t]
	\begin{equation}\label{arraygain}
		\eta\left(f_m\right)=\left|\sum_{k_y=1}^{K_y}\sum_{k_z=1}^{K_z}e^{-j2\pi f_mt_{k_y,k_z}}\sum_{s_y=1}^{S}\sum_{s_z=1}^{S}e^{j2\pi\frac{f_m}{c}\left(r^{k_y,k_z}_{s_y,s_z}+l^{k_y,k_z}_{s_y,s_z}\right)}e^{j\theta^{k_y,k_z}_{s_y,s_z}}\right|.\vspace{-0.4cm}
	\end{equation}\vspace{-0.4cm}
	\end{figure*}
	
	\vspace{-0.2cm}
	\section{Extension to Joint Beamforming for Solving Double-Stage Beam Split Effect}
	In this section, we extend our discussion from single-antenna to multi-antenna at the BS, where beam split occurs at both the IRS and the BS. In fact, the near-field double-stage beam split is coupled, making an optimal beamforming solution unattainable.  Building on a similar approach as in the previous section,  we propose a sub-optimal joint beamforming scheme to address this issue.
	 
	Again, we start by analyzing the distance of between the BS and the IRS.    Following segmentation strategy in the previous section, we divide the entire BS antenna array into $K_t$ sub-arrays, each of which contains $N_k$ adjacent antennas, i.e., $N_t=K_t\times N_k$. Then, we can approximate the distance between the $n_k$-th BS antenna in the $k_t$-th sub-array and the $(s_y,s_z)$-th element of $(k_y,k_z)$-th sub-surface as\begin{align}\label{approx}\vspace{-0.3cm}
			\hspace{-0.2cm}r^{n_k,k_t}_{(s_y,s_z),(k_y,k_z)}\approx&r^{k_t}_{k_y,k_z}-\Delta_{s_y}^{S_y}d\sin\left(\psi^{k_t}_{k_y,k_z}\right)\sin\left(\phi^{k_t}_{k_y,k_z}\right)\nonumber\\
			-\Delta_{s_z}^{S_z}&\cos\left(\psi^{k_t}_{k_y,k_z}\right)-\Delta_{n_k}^{N_k}d\cos\left(\psi^{k_t}_{k_y,k_z}\right),\vspace{-0.3cm}
		\end{align}where $r^{k_t}_{k_y,k_z}$ is the distance between the center of $k_t$-th BS sub-array and the center of $(k_y,k_z)$-th IRS sub-surface, while $\phi^{k_t}_{k_y,k_z}$ and $\psi^{k_t}_{k_y,k_z}$ denote the corresponding azimuth and elevation angles, satisfying
	$r^{k_t}_{k_y,k_z}\!=\!\sqrt{x_{\text{b}}^2\!+\!\left(y_{\text{b}}\!-\!\Delta_{k_y}^{K_y}Sd\right)^2\!\!+\!\left(z_{\text{b}}\!-\!\Delta_{k_t}^{K_t}N_kd\!-\!\Delta_{k_z}^{K_z}Sd\right)^2}$, $
	\sin\!\left(\!\phi^{k_t}_{k_y,k_z}\!\right)\!\!=\!\!\sin\!\left(\!\phi_{k_y,k_z}\!\right)$, $
	\sin\!\left(\!\psi^{k_t}_{k_y,k_z}\!\right)\!\!=\!\!\frac{\left(x_{\text{b}}^2+\left(y_{\text{b}}-\Delta_{k_y}^{K_y}Sd\right)^2\right)^{\frac{1}{2}}}{r^{k_t}_{k_y,k_z}}$, and $
	\cos\!\left(\!\psi^{k_t}_{k_y,k_z}\!\right)\!\!=\!\!\frac{z_{\text{b}}-\Delta_{k_t}^{K_t}N_kd-\Delta_{k_z}^{K_z}Sd}{r^{k_t}_{k_y,k_z}}$. Then, the channel between the BS and the IRS can be approximated based on \eqref{approx}, denoted by $\tilde{\mathbf{G}}_m$.

	Motivated by the decomposability of the far-field double-stage beam
	split proved in \cite{wideband_src}, where the beam split at each stage is addressed by the IRS beamforming and BS beamforming separately, we propose a sub-optimal joint beamforming scheme. Recall that the proposed DLDD IRS beamforming strategy in section III-C aims at solving the phase discrepancy of between the signal reflected by different IRS elements, i.e., the entries in one of the columns in $\tilde{\mathbf{G}}_m$. Thus, we design the BS beamforming to mitigate the phase discrepancy along one of the rows in $\tilde{\mathbf{G}}_m$. The design of BS beamforming is given by\vspace{-0.3cm}\begin{equation}\vspace{-0.1cm}
	\mathbf{w}_m=\mathbf{P}\mathbf{d}_m=\operatorname{blkdiag}\left(\left[\mathbf{p}_1,\cdots,\mathbf{p}_{K_t}\right]\right) e^{j2\pi\frac{f_m}{c}\mathbf{t}},
	\end{equation}where $\mathbf{p}_{k_t}$ is the phase shift at the $k_t$-th sub-array, designed as\vspace{-0.2cm}\begin{equation}
	\mathbf{p}_{k_t}\!\!=\!\!\left[\!e^{\!-j2\pi\frac{f_c}{c}\Delta_{1}^{N_k}d\cos\left(\!\psi^{k_t}_{k_y,k_z}\!\right)}\!,\!\cdots\!,\!e^{\!-j2\pi\frac{f_c}{c}\Delta_{N_k}^{N_k}d\cos\left(\!\psi^{k_t}_{k_y,k_z}\!\right)}\!\right]^T\!\!,
	\end{equation}and $\mathbf{t}$ is the resulting TD over the BS antenna array, designed as\vspace{-0.2cm}\begin{equation}\vspace{-0.1cm}
	\mathbf{t}=\left[0,\tau_{1},\cdots,\sum_{i=1}^{K_t-1}\tau_{i}\right],
	\end{equation}where $\tau_{i}=r^{i}_{1,1}-r^{i+1}_{1,1}$. To this end, we have provided the design of the BS beamforming. Together with the IRS beamforming, such joint beamforming scheme solves the double-stage beam split effect sub-optimally.
	
	
	\vspace{-0.2cm}
	
	\section{Numerical Results}
	In this section, numerical results are presented to illustrate the effectiveness of the proposed DLDD IRS beamforming and sub-optimal joint beamforming schemes.  The center of the IRS is located at $(0,0,0)$ m of a three-dimensional coordinate system. The BS and the user are located at $(0,2,-1)$ m and $(2,-4,-2)$, respectively. It is assumed the IRS elements are separated by half of the wavelength, i.e., $d=\frac{\lambda_c}{2}$. Without otherwise specified, the parameters are set as:  $N_t=32$, $K_t=4$, $N_y=N_z=80$, $K_y=K_z=10$, $f_c=300$ GHz, $B=30$ GHz, $M=128$, $P_m=20$ dBm, and $\sigma_m=-120$ dBm.
	

	
\begin{figure*}[!t]
		\centering
		\subfloat[Normalized array gain versus subcarrier index.\label{array_gain}]{\includegraphics[width=1.7in]{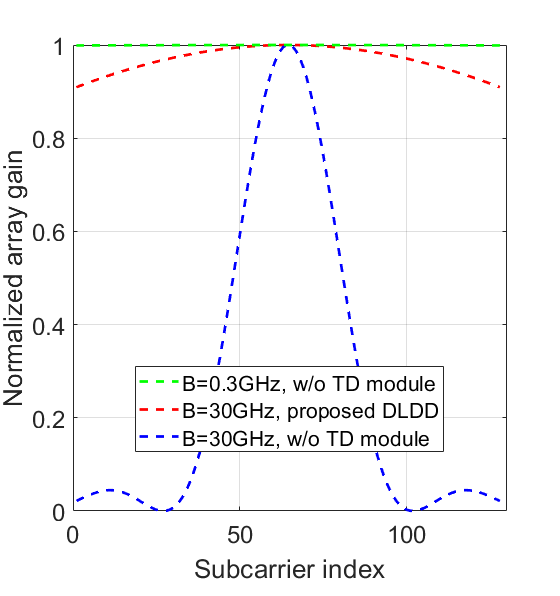}}%
		\hfil
		\subfloat[Normalized edge subcarrier array gain versus $K_T$.\label{arraygain_vs_TTD}]{\includegraphics[width=1.7in]{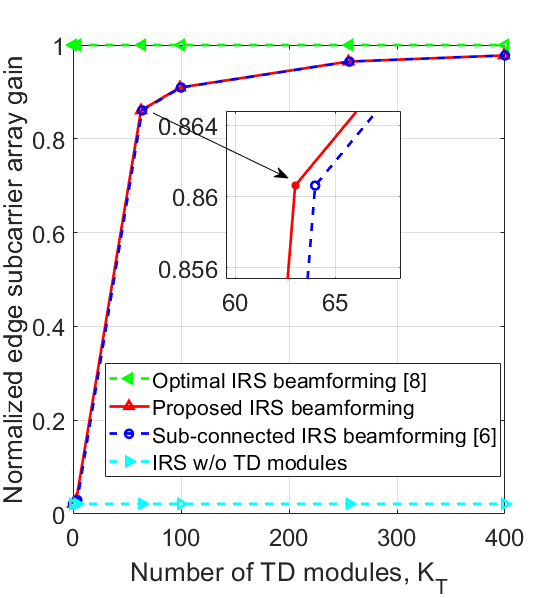}}%
		\hfil
		\subfloat[Normalized edge subcarrier array gain  versus maximum delay range.\label{arraygain_vs_tmax}]{\includegraphics[width=1.7in]{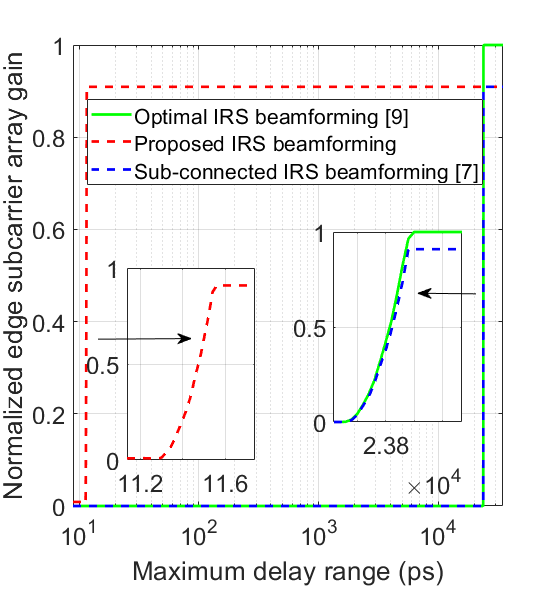}}%
		\hfil
		\subfloat[Average achievable rate per subcarrier versus $B$.\label{rate_pow}]{\includegraphics[width=1.7in]{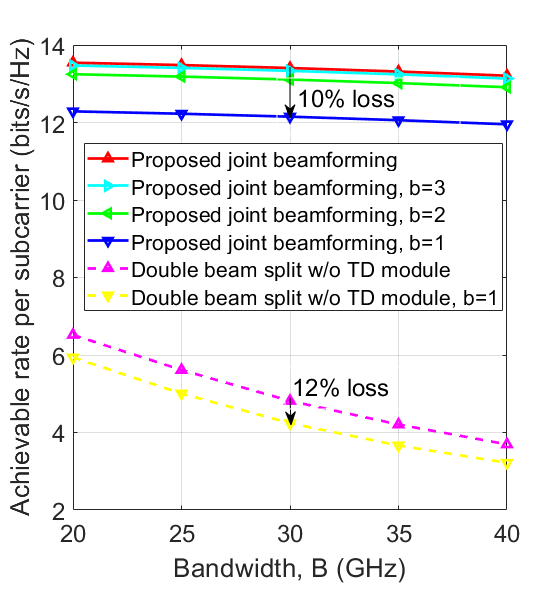}}%

		\caption{Simulation results.}
		\label{best}
		\vspace{-0.4cm}
	\end{figure*}

	
	To validate the effectiveness of the proposed DLDD IRS beamforming scheme, we first considered the single-antenna BS scenario, where $N_t=1$. The generated beams at the center frequency, $f_c$, and the edge subcarriers, $f_1$ and $f_M$, by adopting the proposed DLDD IRS beamforming scheme are presented in Fig. \ref{fig}\subref{proposed_beamsplit}. By comparing Fig. \ref{fig}\subref{beamsplit} and \subref{proposed_beamsplit}, it is observed that the proposed scheme can focus the generated beams at different frequencies to the target location. The normalized array gains at different subcarriers are further illustrated in Fig. \ref{best}\subref{array_gain}. It can be seen that the near-field beam split effect of IRS is suppressed by our proposed scheme. Specifically,  in the wideband system with $B = 30$ GHz, the normalized array gain of conventional narrowband IRS beamforming suffers nearly $100\%$ loss at the edge subcarriers, as compared to the narrowband case with $B = 0.3$ GHz. By performing the proposed  IRS beamforming scheme, the array gain loss can be significantly reduced to about $8\%$. Put differently, the proposed scheme can achieve up to $92\%$ near-optimal array gain across the entire bandwidth. Therefore, it effectively serve as a wideband IRS beamforming design.

	Fig. \ref{best}\subref{arraygain_vs_TTD}  shows the normalized edge subcarrier array gain versus the number of TD modules, $K_T$. As expected, it increases with larger $K_T$. We can see that even a small number of TD modules, e.g., $K_T=63$, readily achieves a near-optimal performance. It is observed that the proposed scheme achieves the same performance as the sub-connected scheme in \cite{time_delay} with one fewer TD module, validating the calculation in Section III-C, where the proposed scheme requires $K\!-\!1$ TD modules, while the sub-connected scheme requires $K$. Although this difference is not significant, the reduced requirement on the range of the TD modules makes the proposed scheme more practical. This is illustrated in Fig. \ref{best}\subref{arraygain_vs_tmax}, the proposed scheme achieves a normalized  edge subcarrier array gain of $0.92$ with a required delay range of only $12$ ps, whereas both the optimal and sub-connected schemes rely on large-range TD modules, as discussed in Section III.
	

	Then, the rate performance in multi-antenna BS scenario is provided, where the double-stage beam split effect is considered. Specifically, the average achievable rate per subcarrier based on our proposed joint beamforming scheme versus the system bandwidth, $B$, is plotted in Fig. \ref{best}\subref{rate_pow}. It is revealed that the achievable rate of the conventional narrowband beamforming decreases rapidly as the bandwidth increases since it cannot address the double beam split effect, whereas the proposed scheme achieves enhanced rate performance.
	This is because the proposed joint beamforming can compensate the array gain loss caused by the double beam split effect even with large bandwidth. The impact of low-resolution IRS phase shifts is also provided by quantizing 
	the obtained continuous phase shifts. It can be observed that the narrowband beamforming scheme suffers more performance loss from the discrete phase quantization than the proposed scheme, which implies the robustness of the latter against the practical hardware limitation. In fact, the $b=2$-bit phase shifters is an efficient choice in practical design, as its performance closely approaches that of continuous phase shifters, and it enjoys a lower complexity and hardware cost.

	
	
	\vspace{-0.2cm}
	\section{Conclusion}
	In this paper, we proposed a novel DLDD architecture that addresses the near-field beam spit effect for the IRS-aided THz communication system. Initially, the near-field beam split effect at the IRS was analyzed, which revealed that the beams at different frequencies point towards different locations. To address this issue, we proposed a DLDD beamforming architecture by extending the piecewise far-field model to the UPA-structured IRS. Then, we also proposed a sub-optimal joint beamforming scheme to address the double-stage beam split at the BS and the IRS. Simulation results demonstrated that the proposed DLDD IRS beamforming scheme can mitigate the near-field beam split effect and achieve a near-optimal performance with lower hardware cost. Besides, the proposed joint beamforming scheme effectively enhanced the rate performance with strong robustness across a wide bandwidth range.
	
	
	
	\vspace{-0.4cm}	
	\bibliographystyle{IEEEtran}
	\bibliography{THz_letter}
\end{document}